\documentclass[aps,prb,twocolumn,superscriptaddress,notitlepage,nobibnotes]{revtex4-1}
\usepackage[colorlinks=true, citecolor=magenta, linkcolor=blue, urlcolor=blue]{hyperref}
\usepackage{graphicx}
\usepackage{amsmath}
\usepackage{amssymb}
\usepackage{amsfonts}
\usepackage{xcolor}
\usepackage{tikz}
\usetikzlibrary{shapes.geometric, arrows}
\usetikzlibrary{decorations.markings}
\newcommand{\bq}{{\bf q}}

\allowdisplaybreaks

\begin{document}
 
\title{Controlling the magnetic state of the proximate quantum spin liquid $\alpha$-RuCl$_3$ with an optical cavity}
\author{Emil Vi\~nas Bostr\"om}
\email{emil.bostrom@mpsd.mpg.de}
\affiliation{Max Planck Institute for the Structure and Dynamics of Matter, Luruper Chaussee 149, 22761 Hamburg, Germany}
\author{Adithya Sriram}
\affiliation{Department of Physics, Stanford University, Stanford, California 94305, USA}
\author{Martin Claassen}
\affiliation{Department of Physics and Astronomy, University of Pennsylvania, Philadelphia, PA 19104}
\affiliation{Center for Computational Quantum Physics, Flatiron Institute, Simons Foundation, New York City, NY 10010, USA}
\author{Angel Rubio}
\email{angel.rubio@mpsd.mpg.de}
\affiliation{Max Planck Institute for the Structure and Dynamics of Matter, Luruper Chaussee 149, 22761 Hamburg, Germany}
\affiliation{Center for Computational Quantum Physics, Flatiron Institute, Simons Foundation, New York City, NY 10010, USA}
\date{\today}

\begin{abstract}
 Harnessing the enhanced light-matter coupling and quantum vacuum fluctuations resulting from mode volume compression in optical cavities is a promising route towards functionalizing quantum materials and realizing exotic states of matter. Here, we extend cavity quantum electrodynamical materials engineering to correlated magnetic systems, by demonstrating that a Fabry-P\'erot cavity can be used to control the magnetic state of the proximate quantum spin liquid $\alpha$-RuCl$_3$. Depending on specific cavity properties such as the mode frequency, photon occupation, and strength of the light-matter coupling, any of the magnetic phases supported by the extended Kitaev model can be stabilized. In particular, in the THz regime, we show that the cavity vacuum fluctuations alone are sufficient to bring $\alpha$-RuCl$_3$ from a zigzag antiferromagnetic to a ferromagnetic state. By external pumping of the cavity in the few photon limit, it is further possible to push the system into the antiferromagnetic Kitaev quantum spin liquid state.
\end{abstract}

\maketitle



\section*{Introduction}
The realization of magnetic van der Waals (vdW) materials with thicknesses down to the monolayer limit has sparked a new interest in fundamental aspects of two-dimensional magnetism.~\cite{Gong17,Burch18,Gong19} Due to a competition of strong anisotropy, fluctuations, and spin-orbit effects, vdW materials are prime candidates to host exotic phenomena such as topological phase transitions, magnetic skyrmions and quantum spin liquids.~\cite{Savary2016,Takagi19} In addition, the electronic, magnetic and optical properties of these materials are sensitive to a wide range of material engineering techniques such as strain,~\cite{Kim2018,Cenker2022} nanostructuring,~\cite{Moll2018} electric fields~\cite{Huang2018,Jiang2018} and moir\'e twisting,~\cite{Xie2021,Kennes2021} allowing their state to be tuned with high precision. 

Recent progress has also established optical engineering techniques as a method to functionalize quantum materials and to reach exotic (out-of-equilibrium) topological phases.~\cite{McIver2019,Rudner2020,Shan2021,Bloch2022} However, driving a system with lasers is associated with excessive heating when the frequency becomes multi-photon resonant with electronic transitions.~\cite{DAlessio2014,Kennes2018} A way to circumvent this problem is to embed the system in an optical cavity, where the effective light-matter coupling is enhanced via mode volume compression and the state of the material can be modified in an equilibrium setting.~\cite{Ruggenthaler2018,Basov2020,Hubener2020,Latini2021,Schlawin2022,Ashida2020,Curtis2022} 
Due to the strong interaction between light and charged excitations polaritonic control of material and chemical properties has, with a few exceptions,~\cite{Sentef2020,Chiocchetta2021,Zhang2022,Dirnberger2022} mainly been considered for electronic states.~\cite{Ebbesen2016,Sidler2021} While currently efforts are made to extend the cavity framework to a broader class of materials, and to construct a unified first principles description of cavity quantum fluctuations and quantum matter,~\cite{Flick2017,Ruggenthaler2018,Hubener2020} experiments demonstrating polaritonic control of materials are scarce.~\cite{Ebbesen2019,Fausti2022} Therefore, to transform this promising approach into a powerful experimental tool, it is of key importance to identify candidate materials where cavity engineering techniques can be explored. 

Here, we extend the concept of cavity quantum electrodynamics (c-QED) engineering into the magnetic regime and identify such a candidate system, by demonstrating how an optical cavity can be used to control the magnetic ground state of the proximate quantum spin liquid $\alpha$-RuCl$_3$ via shaping the quantum fluctuations of the cavity.~\cite{Hubener2020,Ruggenthaler2018,Rokaj2019,Sidler2021} Depending on the cavity frequency, photon occupation, and the strength of the effective light-matter coupling, we find that it is possible to transform the equilibrium zigzag antiferromagnetic order into any of the magnetic phases supported by the extended Kitaev model (see Eq.~\ref{eq:spin_photon_ham} and Fig.~\ref{fig:pgs}). As a key result we find that for frequencies of a few THz and for moderate light-matter couplings, the interaction between the magnetic system and the vacuum fluctuations of the cavity is sufficient to transform $\alpha$-RuCl$_3$ from a zigzag antiferromagnet to a ferromagnet. In contrast to the meta-stable states obtained by driving the system with classical light, the magnetic state resulting from the interaction with the quantum fluctuations of the cavity is a true equilibrium state denoted the photo ground state (PGS).~\cite{Latini2021} Pumping the cavity in the few photon regime, it is further possible to push the system into the Kitaev quantum spin liquid state and to retrieve the non-equilibrium phase diagram of the semi-classical limit.~\cite{Sriram2022} Our results pave the way for utilizing c-QED to induce and control long-lived exotic states in quantum materials.


\begin{figure*}
 \includegraphics[width=\textwidth]{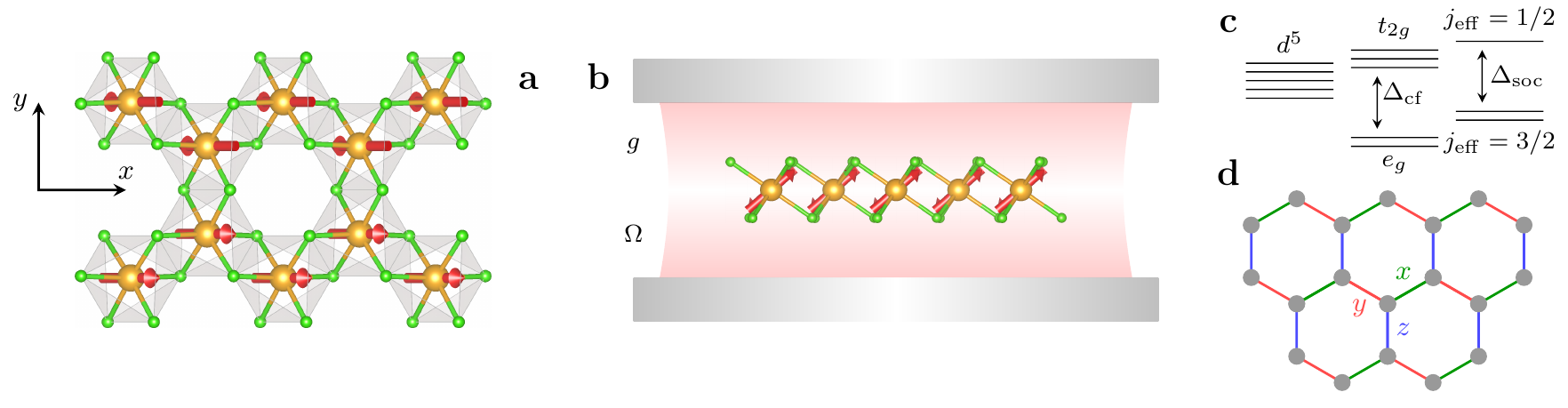}
 \caption{{\bf Crystal and magnetic structure of monolayer $\alpha$-RuCl$_3$.} {\bf a,} Crystal structure of monolayer $\alpha$-RuCl$_3$, with the magnetic Ru ions (orange) arranged in a hexagonal lattice and exhibiting a zigzag antiferromagnetic order. The surrounding octahedra of Cl ions give rise to a crystal field splitting of the Ru $d$-orbitals (see panel {\bf c}.) {\bf b,} The magnetic system interacts with a strength $g$ with a cavity electric field of frequency $\Omega$. {\bf c,} Energy level structure of the magnetic Ru ions, leading to an effective $j_{\rm eff} = 1/2$ magnetic moment in the Ru $t_{2g}$ manifold. {\bf d,} Magnetic interactions of Ru moments, colored according to the Kitaev bonds $x$, $y$ and $z$.}
 \label{fig:structure}
\end{figure*}


\section*{Results}
\subsection*{Low-energy model and equilibrium magnetic order}
The vdW material $\alpha$-RuCl$_3$ consists of layers of Ru atoms arranged in an hexagonal lattice and surrounded by edge-sharing octahedra of Cl ions (Fig.~\ref{fig:structure}a). Due to the crystal field the Ru $d$-orbitals are split by an energy $\Delta_{\rm cf}$ into a lower $e_g$ and a higher $t_{2g}$ manifold, and the strong spin-orbit coupling further splits the $t_{2g}$ states by an energy $\Delta_{\rm soc}$ into a lower $j_{\rm eff} = 3/2$ quartet, fully occupied in the ground state, and a $j_{\rm eff} = 1/2$ doublet with a single hole. The Cl ions are assumed to be completely filled in the ground state, with their $p$-orbitals separated from the Ru $d$-orbitals by a charge-transfer energy $\Delta_{pd}$. The Ru energy level structure is schematically shown in Fig.~\ref{fig:structure}c. The local interactions of the $t_{2g}$ manifold are described by a Hubbard-Kanamori Hamiltonian $H_U$,~\cite{Winter16,Sriram2022} which takes into account the intra-orbital Hubbard interaction $U$, the inter-orbital interaction $U'$, the Hund's coupling $J_H$ and the spin-orbit coupling $\lambda$ (see Methods for a discussion of the model). The Ru holes can either hop directly between Ru atoms or move via the Cl ligands, as described by the kinetic Hamiltonian $H_t$. In the strong coupling limit $t/U \ll 1$ virtual hopping processes give rise to effective magnetic interactions via the superexchange mechanism. Due to the orbital alignment a Kitaev interaction arises from ligand-mediated hopping over $90^\circ$ bond angles,~\cite{Jackeli09} while sub-dominant exchange and anisotropy terms arise from direct Ru-Ru interactions. 

Experimentally $\alpha$-RuCl$_3$ is found to exhibit a zigzag antiferromagnetic order below the N\'eel temperature $T_N \approx 7$ K, as indicated in Fig.~\ref{fig:structure}. At zero temperature, first principles calculations show that this zigzag state is approximately degenerate with a ferromagnetic state,~\cite{Kim2015} and might only be stabilized by spin quantum fluctuations.~\cite{Suzuki2021} In addition, signatures of a Kitaev quantum spin liquid (QSL) state have been found upon applying an external magnetic field along the out-of-plane direction.~\cite{Banerjee2016,Zheng2017} Together these results indicate that although $\alpha$-RuCl$_3$ orders at low temperatures, its ground state is proximate to several competing magnetic orders and the magnetic phase diagram is determined by a delicate competition of different magnetic interactions. This makes $\alpha$-RuCl$_3$ an excellent candidate material to explore the competition between cavity and spin quantum fluctuations. 

In the following the material will be assumed to have $C_3$ symmetry, which is satisfied to a very good degree in $\alpha$-RuCl$_3$. All parameters of the local and kinetic Hamiltonians $H_U$ and $H_t$ were calculated from first principles as discussed in the Methods section, and give a zigzag antiferromagnetic ground state in line with observations.


\subsection*{Light-matter coupling in a cavity}
Within the low-energy description discussed above, the main effect of the cavity is to modify the hopping amplitudes of the Ru holes. Inside the cavity the total Hamiltonian is given by $H = H_U + H_t + \sum_{\lambda} \hbar\Omega_\lambda \hat{n}_\lambda$, where $\hbar\Omega_\lambda$ is the energy of photon mode $\lambda$ and $\hat{n}_\lambda$ is the corresponding number operator. The photon field modifies the kinetic Hamiltonian $H_t$ by introducing the replacements $\hat{c}_{i\alpha\sigma}^\dagger \hat{c}_{i\beta\sigma} \to e^{i \phi_{ij}} \hat{c}_{i\alpha\sigma}^\dagger \hat{c}_{i\beta\sigma}$, where the Peierls phases $\phi_{ij}$ are proportional to the quantum vector potential is $\hat{\bf A} = \sum_{\lambda} ( g_\lambda {\bf e}_\lambda \hat{a}_\lambda^\dagger + g_\lambda^{*} {\bf e}^{*}_\lambda \hat{a}_\lambda)$. For a given mode $\lambda$ the bare light-matter coupling is defined as $g_\lambda = ea/\sqrt{2\epsilon \hbar\Omega_\lambda V}$, where $e$ is the elementary charge, $\hbar$ the reduced Planck constant, $\epsilon$ the relative permittivity and $V$ the cavity mode volume. For a two-dimensional Fabry-P\'erot cavity the lowest energy photon mode has a frequency $\Omega_0 = \pi c/L_z$, and assuming a cavity with linear dimensions $L_x = L_y = L a$ where $L$ is width of the cavity in units of the Ru-Ru distance $a$, the bare light-matter coupling of this mode is $g_0 = e/\sqrt{2\pi\epsilon\hbar c L^2} \approx 0.12/L$. However, below we consider the effective single mode approximation obtained by integrating over photon modes with momenta $\bq \approx 0$, where the effective light-matter coupling $g$ is determined by an effective mode volume $V_{\rm eff}$.~\cite{Choi2017,Hubener2020,Latini2021,Latini2022} Although the value of $V_{\rm eff}$ can in principle be fixed by comparisons to experiment, we here take it as a free parameter.


\begin{figure*}
 \includegraphics[width=\textwidth]{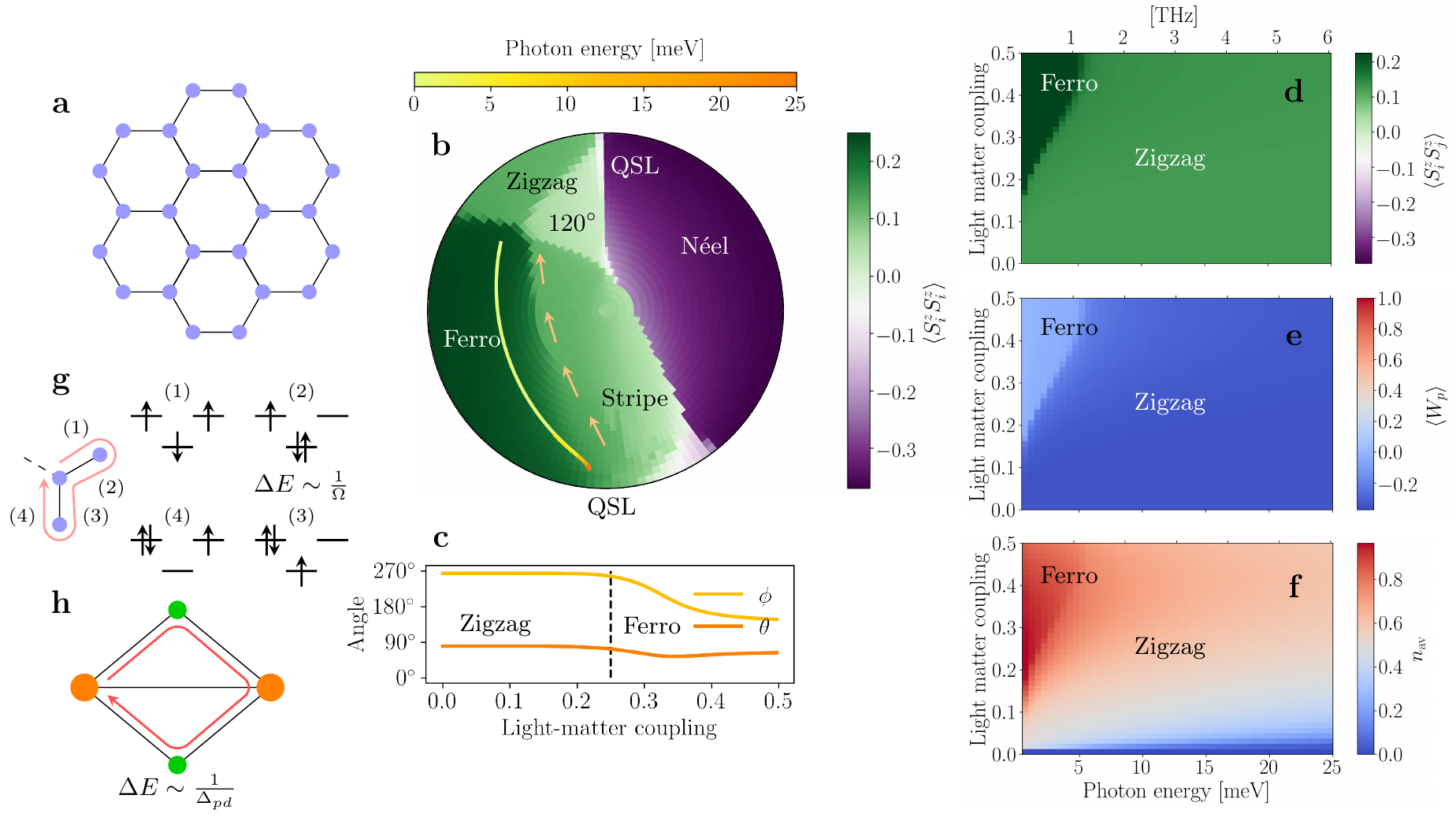}
 \caption{{\bf Magnetic phases of the photo ground state.} {\bf a,} Spin cluster employed for the exact diagonalization studies. {\bf b,} Paths through the magnetic phase diagram traced out by the system as the light-matter coupling $g$ increases from $g_{\rm i} = 0$ to $g_{\rm f} = 0.5$. The paths are colored according to the cavity photon energy $\hbar\Omega$, and the small arrows show the direction of the paths. {\bf c,} Parametrization angles $\theta$ and $\phi$ as a function of $g$ for $\hbar\Omega = 10$ meV. {\bf d,} Nearest neighbor spin-spin correlation functions $\langle S_i^z S_j^z\rangle$. {\bf e,} Expectation value of the plaquette flux operator $W_p$. {\bf f,} Average photon number $n_{\rm av} = \langle \hat{n}\rangle$ of the photo ground state. {\bf g,} Schematic of a fourth order direct process showing a $1/\Omega$ enhancement at low photon energies. {\bf h,} Schematic of a fourth order ligand mediated process unaffected by the low-frequency limit.}
 \label{fig:pgs}
\end{figure*}


\subsection*{Effective coupled spin-photon model}
The total Hamiltonian is down-folded to the spin sector by eliminating $H_t$ to fourth order in virtual ligand mediated processes using quasi-degenerate perturbation theory. The structure of the perturbation expansion allows for the down-folding to be performed separately within each given photon sector, resulting in a coupled spin-photon Hamiltonian of the form
\begin{align}\label{eq:spin_photon_ham}
 \mathcal{H} &= \sum_{\bf nm} \big( \mathcal{H}_{s,\bf nm} + \delta_{\bf nm} \sum_\lambda \hbar\Omega_\lambda n_\lambda \big) |{\bf n}\rangle \langle {\bf m}|.
\end{align}
Here $\mathcal{H}_{s,\bf nm}$ is the spin Hamiltonian in the sector connecting cavity number states $|\bf n\rangle$ and $|\bf m\rangle$, where ${\bf m} = \{m_1, m_2, \ldots, m_L\}$ for $L$ modes. The Hamiltonian in each photon sector is
\begin{align}\label{eq:spin_ham}
  \mathcal{H}_{s,\bf nm} &= \sum_{\langle ij\rangle} \begin{pmatrix} S_i^\alpha & S_i^\beta & S_i^\gamma \end{pmatrix}
  \begin{pmatrix} J & \Gamma & \Gamma' \\
                  \Gamma & J & \Gamma' \\
                  \Gamma' & \Gamma' & J + K \end{pmatrix}_{\bf nm}
  \begin{pmatrix} S_j^\alpha \\ S_j^\beta \\ S_j^\gamma \end{pmatrix}
  \nonumber \\
  &+ B_{\bf nm} \sum_i \hat{\bf e}_B \cdot {\bf S}_i
\end{align}
where the magnetic interactions $J$, $K$, $\Gamma$ and $\Gamma'$ all depend on the light-matter coupling ${\bf g} = \{g_1, g_2, \ldots, g_L\}$ and the photon numbers ${\bf n}$ and ${\bf m}$. Each bond $\langle ij\rangle$ is labeled by the indexes $\alpha\beta(\gamma) \in \{ xy(z), yz(x), zx(y) \}$ and is denoted a $\gamma$-bond in accordance with Fig.~\ref{fig:structure}d. The induced magnetic field $B$ points along the $[111]$ direction of the local spin axes, as defined by the unit vector $\hat{\bf e}_B$.

The form of the spin-photon Hamiltonian in Eq.~\ref{eq:spin_ham} is valid as long as the $C_3$ symmetry is preserved. Since a single linearly polarized mode breaks rotational symmetry, such a mode induces additional terms in the spin Hamiltonian. The rotational symmetry breaking is also reflected in a strong dependence of the cavity-modified spin parameters on the polarization direction, as shown in Supplemental Figure~\ref{fig:linear_cavity}. However, in a Fabry-P\'erot cavity the lowest mode is doubly degenerate with two possible in-plane polarizations, and including both these modes in the description restores the rotational symmetry of $\mathcal{H}$. The $C_3$ symmetry can also be maintained by considering a single circularly polarized mode in a chiral cavity,~\cite{Hubener2020} as will be done in the following.


\begin{figure*} 
 \includegraphics[width=\textwidth]{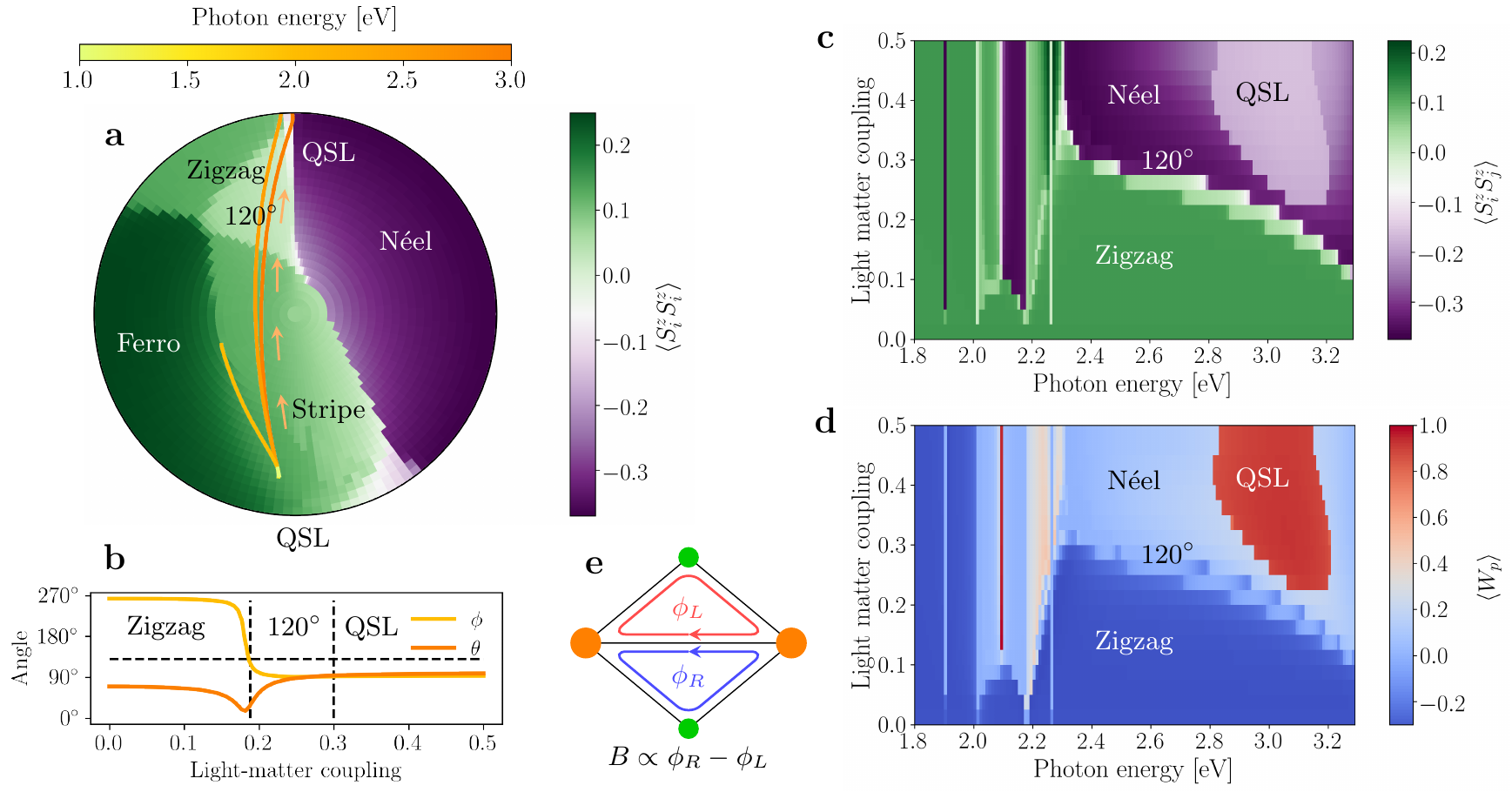}
 \caption{{\bf Magnetic phases in a seeded cavity.} {\bf a,} Paths through the magnetic phase diagram traced out by the spin system as the light-matter coupling $g$ increases from $g_{\rm i} = 0$ to $g_{\rm f} = 0.5$. The paths are colored according to the cavity photon energy $\hbar\Omega$, and the small arrows show the direction of the paths. {\bf b,} Parametrization angles $\theta$ and $\phi$ as a function of $g$ for $\hbar\Omega = 3$ eV. {\bf c,} Spin-spin correlation function $\langle S_i^z S_j^z\rangle$ as a function of photon energy and light-matter coupling. {\bf d,} Average of the plaquette flux operator $W_p$ as a function of photon energy and light-matter coupling. {\bf e,} Schematic of third order processes where hopping around isosceles triangles leads to an induced magnetic field. In all panels the average photon number is $n_{\rm av} = 1$.}
 \label{fig:seeded_phase diagram}
\end{figure*}


\subsection*{Magnetic phases of the cavity photo ground state}
The PGS and magnetic phase diagram of the cavity-embedded spin system was obtained by exact diagonalization of Eq.~\ref{eq:spin_photon_ham} on the 24-site spin cluster shown in Fig.~\ref{fig:pgs}, including a single effective circularly polarized cavity mode. This spin cluster is the minimal one known to respect all sub-lattice symmetries of the magnetic system.~\cite{Chaloupka2015} To perform the calculations we assume a perfect cavity, while a more quantitative description should account for dissipation. However, the results discussed below are expected to be robust against such effects.

To find how the magnetic ground state flows through the phase diagram of the extended Kitaev model as a function of light-matter coupling and cavity frequency (see Fig.~\ref{fig:pgs}b), the magnetic interactions are parameterized by $\bar{J} = \sin\theta \cos\phi$, $\bar{K} = \sin\theta \sin\phi$ and $\bar{\Gamma} = \cos\theta$. Here $\bar{J}$, $\bar{K}$ and $\bar{\Gamma}$ are the cavity-renormalized spin parameters divided by the energy $E = \sqrt{J^2 + K^2 + \Gamma^2}$. The spin parameters for $g = 0$ were obtained via first principles calculations as discussed in the Methods section, and give $\phi = -89.43^\circ$ and $\theta = 69.44^\circ$ placing the equilibrium system in a zigzag antiferromagnetic state adjacent to the ferromagnetic Kitaev QSL. This is in good agreement with experimental findings.~\cite{Banerjee2016,Zheng2017} To identify the magnetic state we calculate the spin-spin correlation function $\langle S_i^z S_j^z \rangle$ and the expectation value of the plaquette flux operator $W_p = 2^{-6} S_i^x S_j^y S_k^z S_l^x S_m^y S_n^z$. In $W_p$ the site indexes traverses a given hexagon $p$ in the positive direction, and the spin component at a given site is determined by the bond pointing out from the hexagon. Magnetic phase transitions are identified by the second order derivatives of the total energy.

Fig.~\ref{fig:pgs}b shows the paths traced out by the magnetic ground state as $g$ is increased from the initial value $g_i = 0$ to the final value $g_f = 0.5$, for a number of photon energies in the range $\hbar\Omega \in [0,0.2]$ eV. For all parameters the system flows away from the ferromagnetic Kitaev point, and depending on $\hbar\Omega$ either remain in the zigzag state or enter the ferromagnetic domain. In particular, for photon energies below $\hbar\Omega \approx 10$ meV, the PGS evolves from a zigzag antiferromagnetic state into a ferromagnetic state as the light-matter coupling is increased (see Figs.~\ref{fig:pgs}d and ~\ref{fig:pgs}e). We remark that light-matter couplings of this magnitude have been achieved in experiments on two-dimensional materials in driven cavities.~\cite{Liu2014,Li2018,ParaviciniBagliani2018} The increase of $g$ is associated with an increase in photon occupation, which diverges in the limit $\Omega \to 0$ (see Fig.~\ref{fig:pgs}f). The reason for this divergence is a class of fourth order virtual electronic processes where a single charge excitation moves while simultaneously emitting/absorbing a virtual cavity photon (see Fig.~\ref{fig:pgs}g). Since the magnetic interaction parameters are inversely proportional to the energy difference between connected virtual states, this leads to a $J \sim 1/\Omega$ enhancement in the low frequency limit. This behavior has been validated in small electron-photon clusters (see Supplementary Figure~\ref{fig:electronic_occupation}), which show good qualitative agreement with the larger-scale spin-photon model simulations.

The microscopic origin of the magnetic phase transition is a simultaneous suppression of the Kitaev interaction and increase of the isotropic exchange, as seen from the normalized spin parameters of the zero photon sector displayed in Supplementary Figure~\ref{fig:normalized_parameters}. In particular, the Kitaev interaction changes sign as a function of light-matter coupling at a point coinciding with the magnetic phase transition. This change in parameters is consistent with a simultaneous suppression of the effective ligand mediated hopping and increase of direct Ru$-$Ru hopping. To see how this comes about we note that while the direct hopping process depicted in Fig.~\ref{fig:pgs}g implies an enhancement of $J \sim 1/\Omega$ at small frequencies, the ligand mediated process shown in Fig.~\ref{fig:pgs}h is proportional to the reciprocal of the charge-transfer energy. Therefore $K \sim 1/\Delta_{pd}$ is largely unaffected by the low frequency limit, leading to a relative enhancement of $J/K$.


\subsection*{Magnetic phase diagram of a pumped cavity}
The results of Fig.~\ref{fig:pgs} show that the quantum fluctuations of the cavity vacuum are sufficient to bring about a change in the magnetic ground state of $\alpha$-RuCl$_3$ in the THz regime. In contrast, for photon energies large compared to the magnetic energy scale ($K/\hbar\Omega \ll 1$), the fluctuations of real photons are strongly suppressed. This restricts the system to a single photon sector, and the Hamiltonian is well approximated by keeping only the diagonal terms $\mathcal{H}_{s,\bf nm} = \mathcal{H}_{s,\bf nn} \delta_{\bf nm}$. In this regime the spin and photon fluctuations therefore decouples, and the magnetic ground state can be obtained from a pure spin model with effective cavity-renormalized parameters.

The magnetic phase diagram of $\alpha$-RuCl$_3$, interacting with a single effective circularly polarized cavity mode with average photon occupation $n_{\rm av} = 1$, is shown in Fig.~\ref{fig:seeded_phase diagram}. The results are displayed as a function of the effective light-matter coupling $\bar{g} = 2g\sqrt{n_{\rm av}}$, and again the cavity is seen to make the magnetic ground state flow towards the antiferromagnetic Kitaev point (Fig.~\ref{fig:seeded_phase diagram}a). The cavity has a drastic effect on the spin parameters as the photon energy is tuned through the charge resonances of the underlying electronic system, as illustrated by the sharp variations of the spin-spin correlation function and plaquette flux around $\hbar\Omega = 2$ eV (see Figs.~\ref{fig:seeded_phase diagram}c and \ref{fig:seeded_phase diagram}d as well as Supplementary Figure~\ref{fig:normalized_seeded_parameters}). For a given light-matter coupling this leads to a series of magnetic phase transitions as a function of frequency, stabilizing quasi-stationary states with either ferromagnetic, antiferromagnetic or spiral order depending on the parameters. 

In addition, there is a large region of light-matter couplings $\bar{g} = 0.2 - 0.6$ and photon energies $\hbar\Omega = 2.8 - 3.2$ eV where the system enters the antiferromagnetic Kitaev QSL state. The presence of a Kitaev QSL has been verified by calculations of the spin-spin correlation function as well as the plaquette flux operator $W_p$, which takes on a quantized value $W_p = \pm 1$ in the QSL phase.~\cite{Takagi19} Although the equilibrium system displays a ferromagnetic Kitaev interaction, the QSL state appears in a region of antiferromagnetic Kitaev interaction implying that the cavity induces a sign reversal of $K$. The QSL is found to be stabilized by the effective magnetic field induced by the cavity (Fig.~\ref{fig:seeded_phase diagram}e), as well as a relative increase of the $\Gamma'$ interaction.

As seen from Fig.~\ref{fig:seeded_phase diagram}, $\alpha$-RuCl$_3$ also enters the QSL phase in a narrow region of cavity mode frequencies around $\hbar\Omega = 2.1$ eV. This coincides with a region where the photo-modified exchange interaction $J$ changes sign, and therefore is very small. More specifically, since the photo-modified magnetic exchange is given close to a resonance by $J \sim 1/(\Omega - E_{\rm res})$, and two resonances appear close together at about $\hbar\Omega = 2.0$ eV and $\hbar\Omega = 2.2$ eV, the different sign of $J$ above the lower and below the upper resonance implies a sign reversal between the resonances. This gives a relative amplification of the Kitaev interaction $K$ inducing the QSL state.


\section*{Discussion}
The results of Fig.~\ref{fig:pgs} and \ref{fig:seeded_phase diagram} show that the magnetic ground state of $\alpha$-RuCl$_3$ is sensitive to the interaction with the quantum fluctuations of the cavity field, and depending on the photon frequency, average cavity occupation and strength of the light-matter coupling, the equilibrium magnetic order can be transformed into any of the magnetic states supported by the extended Kitaev model of Eq.~\ref{eq:spin_photon_ham}. In particular, our results show that the cavity vacuum fluctuations alone are indeed sufficient to change the magnetic order of the photo ground state when the light-matter coupling is sufficiently strong (but within the range of  present experimental light-matter couplings). However, as the value of the light-matter coupling is strongly dependent on the cavity size, and scales as $g \sim 1/L$ with the linear size of the system, it seems as though the effect of the cavity vanishes quickly in the thermodynamic limit. This naive scaling argument might fail for several reasons, the most important being the neglect of additional cavity modes with momenta $\bq \approx 0$. In fact, the number of modes with momenta smaller than some given cut-off $q_{\rm max}$ scales as $N(q_{\rm max}) \sim L$,~\cite{Latini2021,Latini2021b,Fausti2022} and can potentially cancel the volume scaling of $g$. For the effective single mode approximation employed here, the mode volume appearing in the light-matter coupling should therefore be interpreted as an effective mode volume to be fixed either by additional calculations or by comparison to experiment.~\cite{Choi2017,Hubener2020,Latini2021}

An additional subtlety of the thermodynamic limit is the potential coupling of the cavity modes to collective fluctuations of the magnetic system. In particular, close to a magnetic phase boundary, collective fluctuations with a correlation length $\xi \sim L$ are expected to appear. Such collective modes can be expected to play a crucial role in stabilizing the magnetic order, and to couple more strongly to the cavity field.~\cite{Weber2022} However, an explicit treatment of such macroscopic effects, while simultaneously retaining an exact description of spin and photon quantum fluctuations, requires a more advanced methodology going beyond the current work. 

For practical purposes, we note that for a given light-matter coupling $g$ the effective coupling $\bar{g}$ can be enhanced by increasing the photon occupation of the cavity. In particular, since $\bar{g} = 2g\sqrt{n_{\rm av}}$, a substantial effective coupling (and corresponding modification of the magnetic interactions in the spin-photon Hamiltonian of Eq.~\ref{eq:spin_photon_ham}) can be obtained with relatively few photons. Although the present work neglects any cavity dissipation as might be relevant in a non-equilibrium setting or for a dark cavity with lossy mirrors, the qualitative conclusions are expected to be robust against such effects. This includes in particular the identification of a zigzag antiferromagnetic to ferromagnetic transition induced by the cavity vacuum fluctuations, and the presence of a large Kitaev QSL domain in the magnetic phase diagram in the few photon regime. Together these results demonstrate the feasibility of c-QED engineering of the magnetic state of the proximate spin liquid $\alpha$-RuCl$_3$ in the few photon limit. Looking ahead, we note that the cavity photon momentum constitutes a tuning parameter in addition to the light-matter coupling, and we therefore envisage that c-QED can be used to also modify phases with finite-$\bq$ ordering vectors (such as charge density waves). Our work thereby paves the way for utilizing c-QED to control exotic magnetic states in real cavity-embedded quantum materials.~\cite{Ruggenthaler2018,Hubener2020}


\bibliography{references}


\clearpage

\section*{Methods}
\subsection*{Effective low-energy lattice electron Hamiltonian}
The crystal structure of a single layer $\alpha$-RuCl$_3$ consists of a hexagonal lattice of Ru atoms, surrounded by edge-sharing octahedra of Cl atoms. Due to the strong octahedral crystal field the Ru$^{3+}$ $d$-orbitals are split into a lower $e_g$ and a higher $t_{2g}$ manifold, the latter consisting of the orbitals $d_{xy}$, $d_{xz}$ and $d_{yz}$. Due to the strong spin-orbit coupling the $t_{2g}$ states are further split into a lower $j_{\rm eff} = 3/2$ quartet, fully occupied in the ground state, and a $j_{\rm eff} = 1/2$ doublet with a single hole. The Cl$^-$ ions are assumed to be completely filled in the ground state, with their $p_x$, $p_y$ and $p_z$ orbitals separated from the $d$-orbitals by a charge-transfer energy $\Delta_{pd}$.

The local part of the Hamiltonian is given by~\cite{Winter16,Sriram2022}
\begin{align}\label{meth:kanamori}
 H_U &= U \sum_{i\alpha} \hat{n}_{i\alpha\uparrow} \hat{n}_{i\alpha\downarrow} + \sum_{i\sigma\sigma',\alpha<\beta} (U' - J_H \delta_{\sigma\sigma'}) \hat{n}_{i\alpha\sigma} \hat{n}_{i\beta\sigma'} \nonumber \\
 &+ J_H \sum_{i,\alpha\neq\beta} (\hat{c}_{i\alpha\uparrow}^\dagger \hat{c}_{i\alpha\downarrow}^\dagger \hat{c}_{i\beta\downarrow} \hat{c}_{i\beta\uparrow} - \hat{c}_{i\alpha\uparrow}^\dagger \hat{c}_{i\alpha\downarrow} \hat{c}_{i\beta\downarrow}^\dagger \hat{c}_{i\beta\uparrow}) \nonumber \\
 &+ \Delta_{pd} \sum_{i'\sigma} \hat{n}_{i\sigma}^p + \frac{\lambda}{2} \sum_i \hat{\bf c}_i^\dagger ({\bf L} \cdot {\bf s}) \hat{\bf c}_i.
\end{align}
Here $U$ is the intraorbital Hubbard interaction, $U'$ the interorbital interaction and $J_H$ the Hund's coupling between the Ru $d$-orbitals $\alpha,\beta \in \{yz, xz, xy \}$. Further, $\Delta_{pd}$ is the single-particle charge transfer energy to add a hole to the Cl ligands, $\lambda$ is the strength of the spin-orbit coupling (SOC), and the vector of operators is $\hat{\bf c}_i^\dagger = (\hat{c}_{iyz\uparrow}^\dagger, \hat{c}_{iyz\downarrow}^\dagger, \hat{c}_{ixz\uparrow}^\dagger, \hat{c}_{ixz\downarrow}^\dagger, \hat{c}_{ixy\uparrow}^\dagger, \hat{c}_{ixy\downarrow}^\dagger)$. In this basis, the inner product of orbital and spin angular momentum may be written
\begin{align}
 {\bf L} \cdot {\bf s} = \begin{pmatrix} 0 & -i\sigma_z & i\sigma_y \\
                                         i\sigma_z & 0 & -i\sigma_x \\
                                         -i\sigma_y & i\sigma_x & 0 \end{pmatrix}.
\end{align}

To account for hopping processes between the Ru $d$-orbitals and Cl $p$-orbitals, we assume the system has $C_3$ symmetry. The hopping processes along a $Z$-bond are described by
\begin{align}\label{meth:hopping}
 H_t' &= \sum_{\langle ij\rangle\sigma} \bigg[
 \Big( \hat{c}_{iyz\sigma}^\dagger \, \hat{c}_{ixz\sigma}^\dagger \, \hat{c}_{ixy\sigma}^\dagger \Big)
 \begin{pmatrix} t_1 & t_2 & t_4 \\ t_2 & t_1 & t_4 \\ t_4 & t_4 & t_3 \end{pmatrix}
 \begin{pmatrix} \hat{c}_{iyz\sigma} \\ \hat{c}_{ixz\sigma} \\ \hat{c}_{ixy\sigma} \end{pmatrix} \nonumber \\
 &+ t_{pd} \big( \hat{p}_{i'\sigma}^\dagger \hat{c}_{ixz\sigma} + \hat{c}_{jyz\sigma}^\dagger \hat{p}_{i'\sigma} \nonumber \\
 &\hspace*{0.55cm}+ \hat{p}_{j'\sigma}^\dagger \hat{c}_{iyz\sigma} + \hat{c}_{jxz\sigma}^\dagger \hat{p}_{j'\sigma} \big) + H.c. \bigg].
\end{align}
Here, $t_i$ with $i \in \{1,2,3,4\}$ parameterize direct $d-d$ hopping processes, while $t_{pd}$ determines the strength of ligand mediated hopping via the Cl $p_z$-orbitals.


\subsection*{Light-matter coupling}
In presence of the cavity the total Hamiltonian in the dipole approximation is given by $H = H_U + H_t + \sum_{\lambda} \hbar\Omega_\lambda \hat{n}_\lambda$. Here, $\hbar\Omega_\lambda$ is the photon energy of a mode described by the label $\lambda$ and $\hat{n}_\lambda$ is the corresponding number operator. Inside the cavity the Hamiltonian $H_t$ is modified by the replacements $\hat{c}_{i\alpha\sigma}^\dagger \hat{c}_{i\beta\sigma} \to e^{i \phi_{ij}} \hat{c}_{i\alpha\sigma}^\dagger \hat{c}_{i\beta\sigma}$, where the Peierls phases are $\phi_{ij} = (ea/\hbar) {\bf d}_{ij} \cdot \hat{\bf A} $, ${\bf d}_{ij} = {\bf r}_j - {\bf r}_i$ is the vector between atomic sites $i$ and $j$ measured in units of the Ru-Ru distance $a$, and the quantum vector potential is
\begin{align}
 \hat{\bf A} = \sum_{\lambda} ( A_\lambda {\bf e}_\lambda \hat{a}_\lambda^\dagger + A_\lambda^{*} {\bf e}^{*}_\lambda \hat{a}_\lambda).
\end{align}
For a given mode $\lambda$ this defines the light-matter coupling $g_\lambda = (ea/\hbar) A_\lambda = ea/\sqrt{2\epsilon_0 \hbar\Omega_\lambda V}$. Thus, the Peierls phases can be written as $\phi_{ij} = {\bf d}_{ij} \cdot \hat{\bf A}$ with
\begin{align}
 \hat{\bf A} = \sum_{\lambda} ( g_\lambda {\bf e}_\lambda \hat{a}_\lambda^\dagger + g_\lambda^{*} {\bf e}^{*}_\lambda \hat{a}_\lambda),
\end{align}
and the modified hopping Hamiltonian is
\begin{align}
 &H_t = \sum_{\langle ij\rangle\sigma} \bigg[
 \Big( \hat{c}_{iyz\sigma}^\dagger \, \hat{c}_{ixz\sigma}^\dagger \, \hat{c}_{ixy\sigma}^\dagger \Big)
 e^{i{\bf d}_{ij} \cdot \hat{\bf A}} \begin{pmatrix} t_1 & t_2 & t_4 \\ t_2 & t_1 & t_4 \\ t_4 & t_4 & t_3 \end{pmatrix}
 \begin{pmatrix} \hat{c}_{iyz\sigma} \\ \hat{c}_{ixz\sigma} \\ \hat{c}_{ixy\sigma} \end{pmatrix} \nonumber \\
 &+ t_{pd} \big( e^{i{\bf d}_{i'i} \cdot \hat{\bf A}} \hat{p}_{i'\sigma}^\dagger \hat{c}_{ixz\sigma} + e^{i{\bf d}_{ji'} \cdot \hat{\bf A}} \hat{c}_{jyz\sigma}^\dagger \hat{p}_{i'\sigma} \nonumber \\
 &\hspace*{0.55cm}+ e^{i{\bf d}_{j'i} \cdot \hat{\bf A}} \hat{p}_{j'\sigma}^\dagger \hat{c}_{iyz\sigma} + e^{i{\bf d}_{jj'} \cdot \hat{\bf A}} \hat{c}_{jxz\sigma}^\dagger \hat{p}_{j'\sigma} \big) + H.c. \bigg]. 
\end{align}


\subsection*{Hamiltonian in the photon number basis}
The Hamiltonian is expanded in the photon number basis $|{\bf n}\rangle = |n_1, n_2, \ldots, n_N \rangle$ according to~\cite{Sentef2020,Li2020,Li2021}
\begin{align}
 H &= \sum_{\bf nm} ({\bf 1}_e \otimes |{\bf n}\rangle \langle {\bf n}|) H ({\bf 1}_e \otimes |{\bf m}\rangle \langle {\bf m}|) \nonumber \\
 &= \sum_{\bf nm} H_{\bf nm} \otimes |{\bf n}\rangle \langle {\bf m}|.
\end{align}
Here ${\bf 1}_e$ is the identity operator in the electronic Hilbert space, and the Hamiltonian $H_{\bf nm}$ is given by
\begin{align}
 H_{\bf nm} &= \big( H_U + \sum_\lambda \hbar\Omega_\lambda n_\lambda \big) \delta_{\bf nm} + H_{t,\bf nm}.
\end{align}
The matrix elements $\langle {\bf n}| e^{i{\bf d}_{ij} \cdot \hat{\bf A}} |{\bf m}\rangle$ are obtained by noting that since $[\hat{a}_\lambda^\dagger, \hat{a}_{\lambda'}] = 0$, the Peierls phases factorize over different modes and
\begin{align}
 \langle {\bf n}| e^{i{\bf d}_{ij} \cdot \hat{\bf A}} |{\bf m}\rangle = \prod_\lambda \langle n_\lambda| e^{i{\bf d}_{ij} \cdot \hat{\bf A}_\lambda} |m_\lambda\rangle,
\end{align}
where $\hat{\bf A}_\lambda = g_\lambda {\bf e}_\lambda \hat{a}_\lambda^\dagger + g_\lambda^{*} {\bf e}^{*}_\lambda \hat{a}_\lambda$.

The single-mode expressions are calculated by introducing the variables $\eta_{ij\lambda} = g_\lambda ({\bf d}_{ij} \cdot {\bf e}_\lambda)$. Using the Baker-Hausdorff formula to expand the exponential, the matrix elements are $\langle n_\lambda| e^{i{\bf d}_{ij} \cdot \hat{\bf A}_\lambda} |m_\lambda\rangle = i^{|n_\lambda - m_\lambda|} j_{n_\lambda,m_\lambda}^{ij}$ where 
\begin{align}
 j_{n_\lambda,m_\lambda}^{ij} &= e^{-|\eta_{ij\lambda}|^2/2} \hspace*{3.65cm}\text{if}\hspace*{0.2cm} n_\lambda \geq m_\lambda \nonumber \\
 \times& \sum_{k=0}^{m_\lambda} \frac{(-1)^k |\eta_{ij\lambda}|^{2k} (\eta_{ij\lambda})^{n_\lambda-m_\lambda}}{k!(k+n_\lambda-m_\lambda)!} \sqrt{\frac{n_\lambda!}{m_\lambda!}} \frac{m_\lambda!}{(m_\lambda-k)!} \\
 j_{n_\lambda,m_\lambda}^{ij} &= e^{-|\eta_{ij\lambda}|^2/2} \hspace*{3.65cm}\text{if}\hspace*{0.2cm} m_\lambda \geq n_\lambda \nonumber \\
 &\times \sum_{k=0}^{n_\lambda} \frac{(-1)^k |\eta_{ij\lambda}|^{2k} (\eta_{ij\lambda}^{*})^{m_\lambda-n_\lambda}}{k!(k+m_\lambda-n_\lambda)!} \sqrt{\frac{m_\lambda!}{n_\lambda!}} \frac{n_\lambda!}{(n_\lambda-k)!}.
\end{align}
Using the notation ${\bf g} = \{ g_1, g_2, \ldots, g_N \}$ to denote the full set of light-matter couplings, the light-matter coupling is described by the function
\begin{align}
 J_{\bf nm}^{ij}({\bf g}) = \langle {\bf n}| e^{i{\bf d}_{ij} \cdot \hat{\bf A}} |{\bf m}\rangle = \prod_\lambda j_{n_\lambda,m_\lambda}^{ij}.
\end{align}
With this function, the hopping Hamiltonian in presence of the cavity can be written as
\begin{align}\label{meth:hopping_photon}
 H_{t,\bf nm} &= \sum_{\langle ij\rangle\sigma} \bigg[
 \Big( \hat{c}_{iyz\sigma}^\dagger \, \hat{c}_{ixz\sigma}^\dagger \, \hat{c}_{ixy\sigma}^\dagger \Big) \nonumber \\
 &\times J_{\bf nm}^{ij}({\bf g}) \begin{pmatrix} t_1 & t_2 & t_4 \\ t_2 & t_1 & t_4 \\ t_4 & t_4 & t_3 \end{pmatrix}
 \begin{pmatrix} \hat{c}_{iyz\sigma} \\ \hat{c}_{ixz\sigma} \\ \hat{c}_{ixy\sigma} \end{pmatrix} \nonumber \\
 &+ t_{pd} \big( J_{\bf nm}^{i'i}({\bf g}) \hat{p}_{i'\sigma}^\dagger \hat{c}_{ixz\sigma} + J_{\bf nm}^{ji'}({\bf g}) \hat{c}_{jyz\sigma}^\dagger \hat{p}_{i'\sigma} \\
 &\hspace*{0.55cm}+ J_{\bf nm}^{j'i}({\bf g}) \hat{p}_{j'\sigma}^\dagger \hat{c}_{iyz\sigma} + J_{\bf nm}^{jj'}({\bf g}) \hat{c}_{jxz\sigma}^\dagger \hat{p}_{j'\sigma} \big) + H.c. \bigg]. \nonumber
\end{align}


\subsection*{Single-mode light-matter coupling and polarization dependence}
For a single photon mode the index $\lambda$ can be dropped, and the function $J_{\bf nm}^{ij}({\bf g}) = J_{nm}^{ij}(g) = j_{n,m}^{ij}$. Writing the intersite vector on a given bond as ${\bf d}_{ij} = d(\cos\theta_{ij}\hat{\bf x} + \sin\theta_{ij}\hat{\bf y})$, the parameter $\eta_{ij}$ is given for a circularly polarized field by $\eta_{ij} = g ({\bf d}_{ij} \cdot {\bf e}) = gd e^{\pm i\theta_{ij}}$, where $\pm$ denotes a left-handed/right-handed polarization. Noting that $|\eta_{ij}| = |\eta| = gd$ so that the spatial dependence of the function $j_{n,m}^{ij}$ factorizes, we have $j_{n,m}^{ij} = j_{n,m} e^{\pm i(n-m)\theta_{ij}}$ with
\begin{align}
 j_{n,m} &= e^{-|\eta|^2/2} \sum_{k=0}^m \frac{|\eta|^{2k+n-m}}{k!(k+n-m)!} \sqrt{\frac{n!}{m!}} \frac{m!}{(m-k)!}
\end{align}
for $n \geq m$ and similarly but with $n \leftrightarrow m$ for $m \geq n$. 

For linear polarization the polarization vector can be written as ${\bf e} = \cos\phi \hat{\bf x} + \sin\phi \hat{\bf y}$, and we have $\eta_{ij} = gd \cos(\theta_{ij} - \phi)$.


\subsection*{Connection to Floquet Hamiltonian}
As a consistency check, the Hamiltonian $H_{t,\bf nm}$ for a single circularly polarized photon mode is shown to reduce to the Floquet Hamiltonian of Ref.~\cite{Sriram2022} in the semi-classical limit. For a single photon mode, we can drop the mode index, and the function $J_{\bf nm}^{ij}({\bf g}) = J_{nm}^{ij}(g) = j_{n,m}^{ij}$. Taking the polarization vector of the photon mode to be ${\bf e} = (\hat{\bf x} \pm i \hat{\bf y})/\sqrt{2}$, and writing ${\bf d}_{ij} = d(\cos\theta_{ij}\hat{\bf x} + \sin\theta_{ij}\hat{\bf y})$, we have $\eta_{ij} = g ({\bf d}_{ij} \cdot {\bf e}) = gd e^{\pm i\theta_{ij}}$.

To establish the connection to the Floquet Hamiltonian, what remains is now to show that the function $j_{n,m}(g)$ reduces to the Bessel function $J_{n-m}(A)$ in the semi-classical limit. To do this, we write $n = m + l$ and take the limit $m \to \infty$ and $g \to 0$ while keeping $A = 2gd\sqrt{m}$ constant. Noting that
\begin{align}
 j_{m+l,m} &= e^{-|\eta|^2/2} \sum_{k=0}^m \frac{|\eta|^{2k+l}}{k!(k+l)!} \sqrt{\frac{(m+l)!}{m!}} \frac{m!}{(m-k)!} \nonumber \\
 &= e^{-A^2/8n} \sum_{k=0}^m \frac{\left(\frac{A}{2}\right)^{2k+l}}{k!(k+l)!} \sqrt{\frac{(m+l)!}{m!m^l}} \frac{m!}{(m-k)!m^k},
\end{align}
and that $(m+l)!/m! \to m^l$ and $m/(m-k)! \to m^k$ as $m \to \infty$, we have
\begin{align}
 \lim_{m\to\infty} j_{m+l,m} &= \sum_{k=0}^\infty \frac{(-1)^k \left(\frac{A}{2}\right)^{2k+l}}{k!(k+l)!} = J_l(A).
\end{align}
Since $l = n-m$ the Hamiltonian $H_{t,\bf nm}$ reproduces the correct Floquet Hamiltonian in the semi-classical limit.


\subsection*{Cavity-modified spin Hamiltonian}
The total Hamiltonian $H$ is down-folded to the spin sector by using quasi-degenerate perturbation theory to eliminate $H_t$ order by order.~\cite{Winkler2003} This perturbation expansion is performed numerically up to fourth order in $t/U$ using exact diagonalization of the Ru$_2$Cl$_2$ cluster shown in Fig.~\ref{fig:seeded_phase diagram}e. This allows to extract the nearest-neighbor magnetic interactions and their dependence on the light-matter coupling, cavity frequency and polarization. The structure of the perturbation expansion allows the down-folding to be performed separately within each photon sector, giving a coupled spin-photon Hamiltonian of the form
\begin{align}
 \mathcal{H} &= \sum_{\bf nm} \big( \mathcal{H}_{s,\bf nm} + \delta_{\bf nm} \sum_\lambda \hbar\Omega_\lambda n_\lambda \big) |{\bf n}\rangle \langle {\bf m}|.
\end{align}
Here $\mathcal{H}_{s,\bf nm}$ is the spin Hamiltonian in the photon sector connecting photon numbers $\bf n$ and $\bf m$, and is given by
\begin{align}
  \mathcal{H}_{s,\bf nm} &= \sum_{\langle ij\rangle} \begin{pmatrix} S_i^\alpha & S_i^\beta & S_i^\gamma \end{pmatrix}
  \begin{pmatrix} J & \Gamma & \Gamma' \\
                  \Gamma & J & \Gamma' \\
                  \Gamma' & \Gamma' & J + K \end{pmatrix}_{\bf nm}
  \begin{pmatrix} S_j^\alpha \\ S_j^\beta \\ S_j^\gamma \end{pmatrix}
  \nonumber \\
  &+ B_{\bf nm} \sum_i \hat{\bf e}_B \cdot {\bf S}_i
\end{align}
where each bond $(ij)$ is labeled by the indexes $\alpha\beta(\gamma) \in \{ xy(z), yz(x), zx(y) \}$. In this Hamiltonian, the magnetic parameters $J$, $K$ and $\Gamma$ all depend on the light-matter coupling ${\bf g}$ and the photon numbers ${\bf n}$ and ${\bf m}$.


\subsection*{Exact diagonalization}
The ground state of the coupled spin-photon system was obtained by exact diagonalization of a 24 site spin cluster interacting with a single photon mode. To perform the calculations we used to open-source Python package QuSpin.~\cite{Weinberg2017,Weinberg2019}


\begin{figure}
    \centering
    \includegraphics[width=0.9\columnwidth]{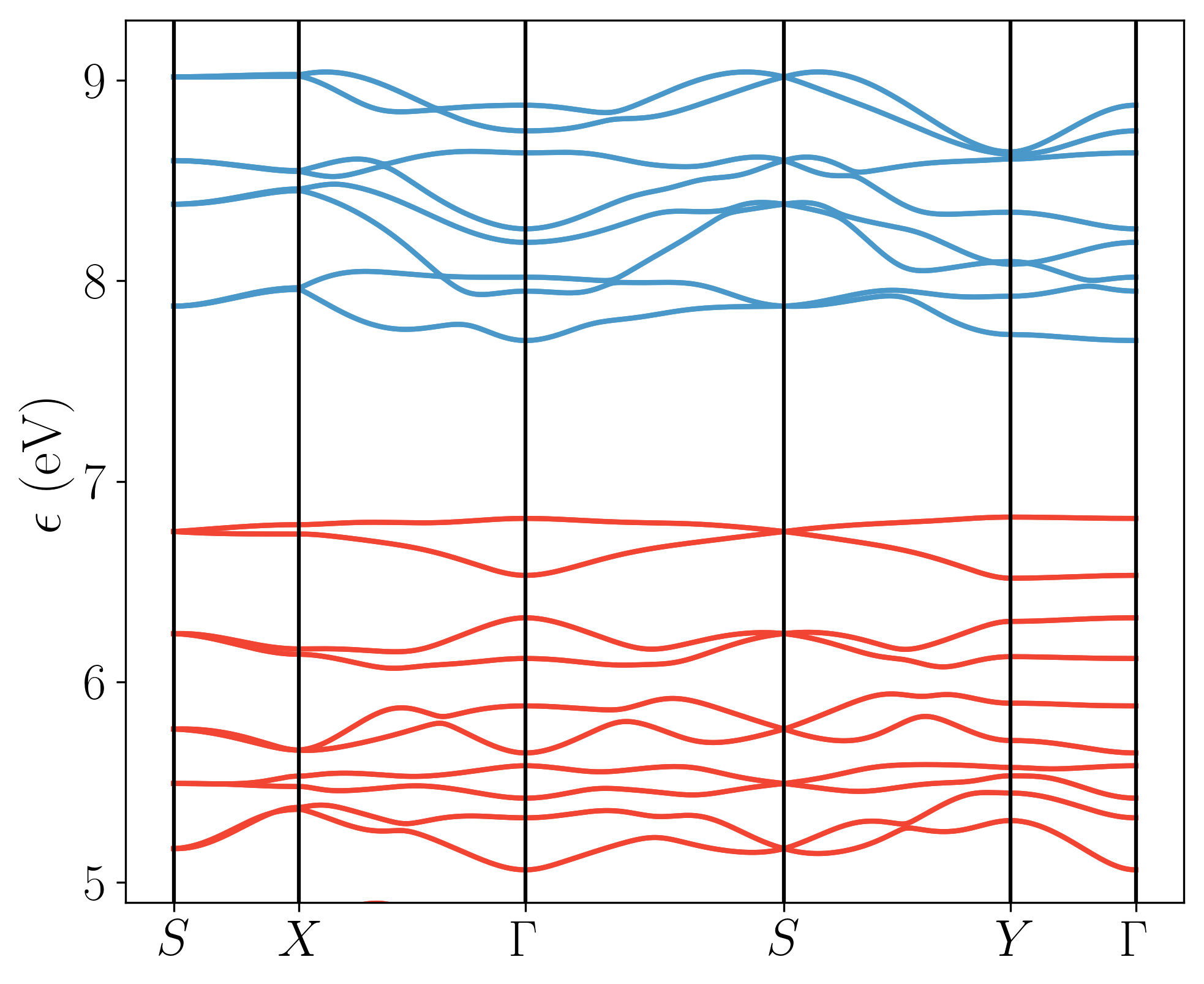}
    \caption{{\bf Electronic bandstructure of $\alpha$-RuCl$_3$.} Highest valence bands (red) and lowest conductions bands (blue) of $\alpha$-RuCl$_3$ obtained within the DFT+$U$ formalism.~\cite{TancogneDejean2017}}
    \label{fig:bandstructure}
\end{figure}

\subsection*{Model parameters from first principles}
To determine the parameters of the electronic Hamiltonians $H_U$ and $H_t$, we performed first principles simulations of monolayer $\alpha$-RuCl$_3$ with the {\sc Octopus} electronic structure code.~\cite{TancogneDejean2017,TancogneDejean2020} The single particle parameters where obtained from a Wannierization of the Ru $4d$ and Cl $3p$ orbitals in the paramagnetic state using Wannier90,~\cite{Marzari97,Souza01} while the interaction parameters where determined using the hybrid DFT+$U$ functional ACBN0 in the zigzag state.~\cite{TancogneDejean2017} We have checked that the single particle parameters differ by less than one percent between the paramagnetic and zigzag states. The resulting electronic parameters are given in Tab.~\ref{tab:parameters}, the equilibrium spin parameters in Tab.~\ref{tab:spin_parameters}, and the electronic band structure is shown in Fig.~\ref{fig:bandstructure}.

The calculations were performed in a $1\times \sqrt{3}$ supercell to account for the zigzag magnetic structure, using the experimental lattice parameters $a = 5.98$ {\AA } and $b = 10.35$ \AA. Mixed boundary conditions, periodic in the in-plane direction and open in the out-of-plane direction, where used together with a vacuum region of $15$ {\AA } to ensure convergence in the out-of-plane direction. A $8\times 8$ $k$-point grid and a real-space grid spacing of $0.3$ Bohr were employed. Using the ACBN0 functional a self-consistent effective interaction $U_{\rm eff} = U - J_H$ was determined on both the Ru and Cl ions, and the Kanamori parameters $U$ and $J_H$ where calculated in the final states after convergence had been reached.

\begin{table}
 \begin{tabular}{c|ccccccccc}
 Parameter & $U$ & $J_H$ & $\lambda$ & $\Delta_{pd}$ & $t_1$ & $t_2$ & $t_3$ & $t_4$ & $t_{pd}$ \\ \hline
 eV & 3.66 & 0.64 & 0.18 & 4.34 & 0.035 & 0.035 & -0.06 & -0.023 & -0.8
 \end{tabular}
 \caption{{\bf Electronic parameters of $\alpha$-RuCl$_3$:} Parameters of the Hubbard-Kanamori Hamiltonian $H_U$ and the hopping Hamiltonian $H_t$ as calculated from first principles with the {\sc Octopus} and {\sc Wannier90} electronic structure codes.}
 \label{tab:parameters}
\end{table}

\begin{table}
 \begin{tabular}{c|cccc}
 Parameter & $J$ & $K$ & $\Gamma$ & $\Gamma'$  \\ \hline
 meV & -0.078 & -7.91 & 2.96 & -0.90 
 \end{tabular}
 \caption{{\bf Equilibrium spin parameters of $\alpha$-RuCl$_3$:} Equilibrium spin parameters of $\alpha$-RuCl$_3$ for the model $\mathcal{H}_s$, calculated from the electronic parameters of Table~\ref{tab:parameters}.}
 \label{tab:spin_parameters}
\end{table}

For a two-dimensional cavity the lowest energy photon mode has a frequency $\Omega = \pi c/L_z$, where $L_z$ is the extent of the cavity in the $z$-direction. Therefore, the light-matter coupling is independent of the photon frequency, and is given by
\begin{align}
 g = \frac{ea}{\sqrt{2\epsilon_0 \hbar\Omega a_x a_y L_z}} = \frac{ea}{\sqrt{2\pi\epsilon_0 \hbar c a_x a_y}} = 0.12
\end{align}
assuming $a_x a_y = a^2$.


\section*{Data Availability}
All data supporting the findings of this study are available from the corresponding authors upon reasonable request.

\section*{Code Availability}
The codes used to generate the data of this study are available from the corresponding authors upon reasonable request.

\section*{Acknowledgements} 
We acknowledge support by the Max Planck Institute New York City Center for Non-Equilibrium Quantum Phenomena, the Cluster of Excellence 'CUI: Advanced Imaging of Matter'- EXC 2056 - project ID 390715994 and  SFB-925 "Light induced dynamics and control of correlated quantum systems" – project 170620586 of the Deutsche Forschungsgemeinschaft (DFG), and Grupos Consolidados (IT1453-22). The Flatiron Institute is a Division of the Simons Foundation.

\section*{Author Contributions}

\section*{Competing Interests}
The authors declare no competing financial or non-financial interests.


\clearpage
\appendix
\renewcommand\thefigure{S\arabic{figure}}
\setcounter{figure}{0}
\widetext



\begin{figure*}
 \includegraphics[width=\textwidth]{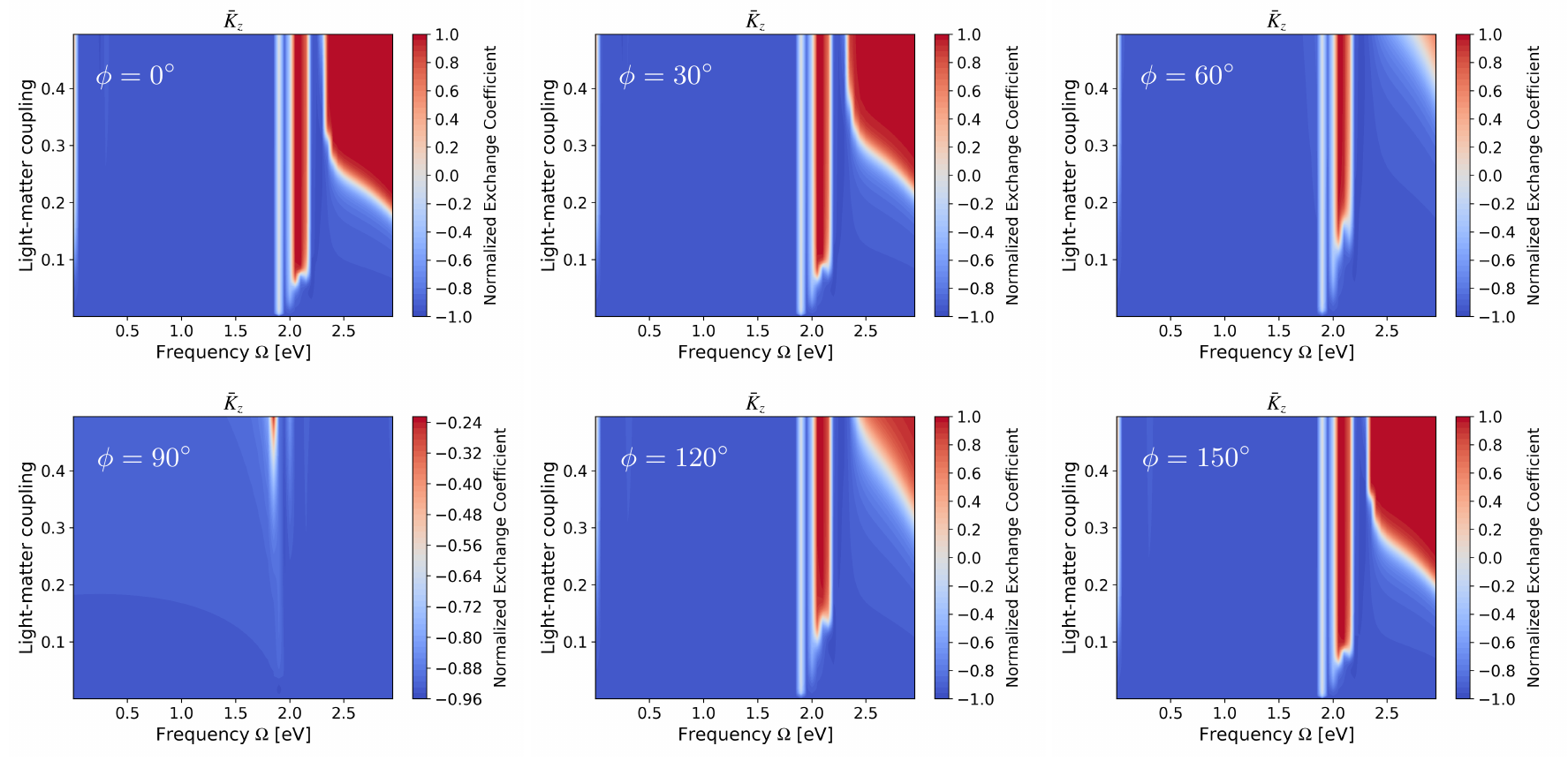}
 \caption{{\bf Polarization dependence of the Kitaev interaction for a linearly polarized cavity mode.} Effective Kitaev interaction $\bar{K}$ as a function of the effective light-matter coupling $\bar{g} = 2g\sqrt{n_{\rm av}}$ and frequency $\Omega$ for a cavity with $n_{\rm av} = 1$. The different panels correspond to polarization vectors making and angle $\phi$ with the $x$-axis. Due to the $C_3$ symmetry of the magnetic system, the polarization dependence shows a period of $\phi_0 = 2\pi/3$.}
 \label{fig:linear_cavity}
\end{figure*}



\begin{figure*}
 \includegraphics[width=\textwidth]{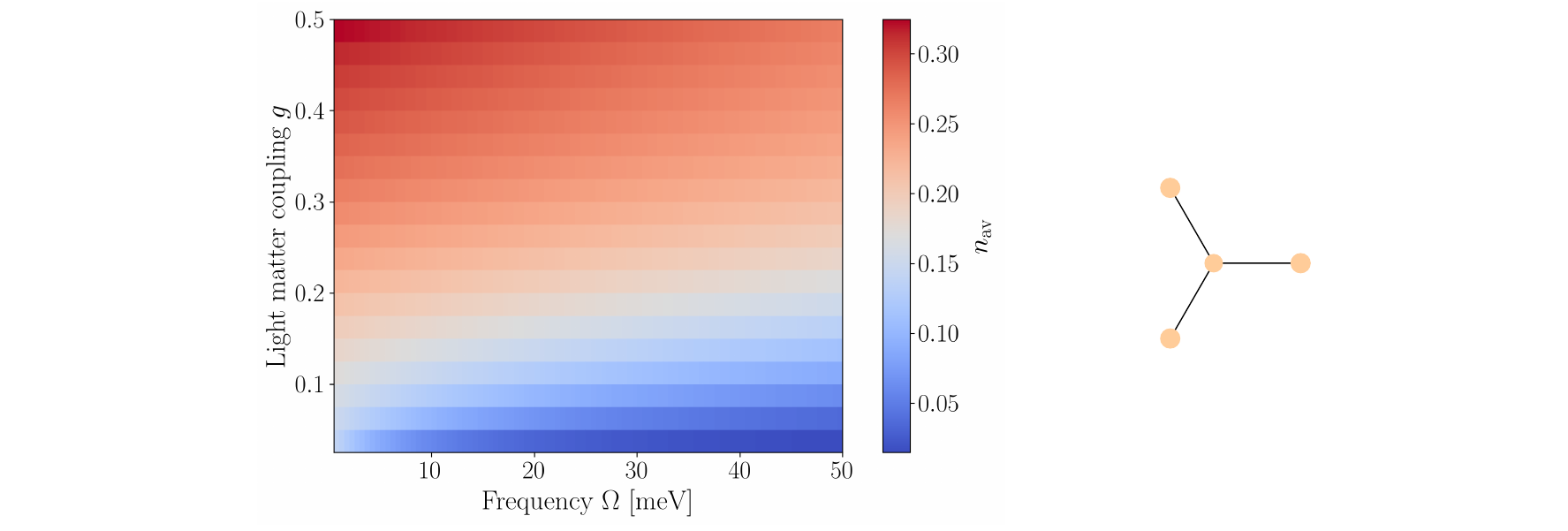}
 \caption{{\bf Photon occupation of an electronic cluster.} Photon occupation of the ground state of a four site Ru cluster described by the Hamiltonian in Eqs.~\ref{meth:kanamori} and \ref{meth:hopping_photon}. The Ru cluster is schematically shown to the right.}
 \label{fig:electronic_occupation}
\end{figure*}



\begin{figure*}
 \includegraphics[width=\textwidth]{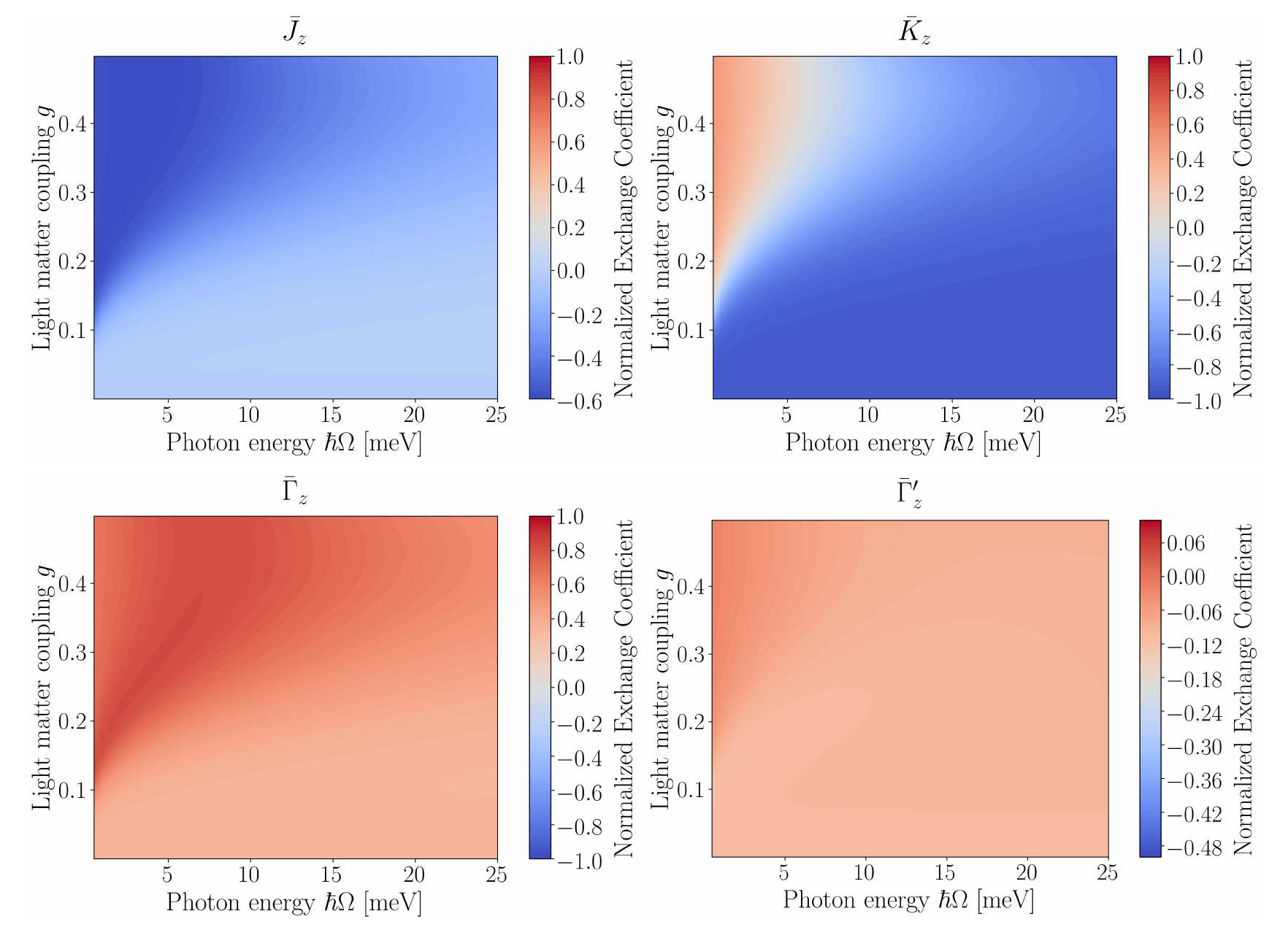}
 \caption{{\bf Spin parameters of the dark cavity.} Normalized magnetic exchange interaction $\bar{J}$, Kitaev interaction $\bar{K}$, and anisotropy interactions $\bar{\Gamma}$ and $\bar{\Gamma}'$ as a function of the light-matter coupling $g$ and the photon energy $\hbar\Omega$ in the zero photon sector $n_{\rm av} = 0$.}
 \label{fig:normalized_parameters}
\end{figure*}



\begin{figure*}
 \includegraphics[width=\textwidth]{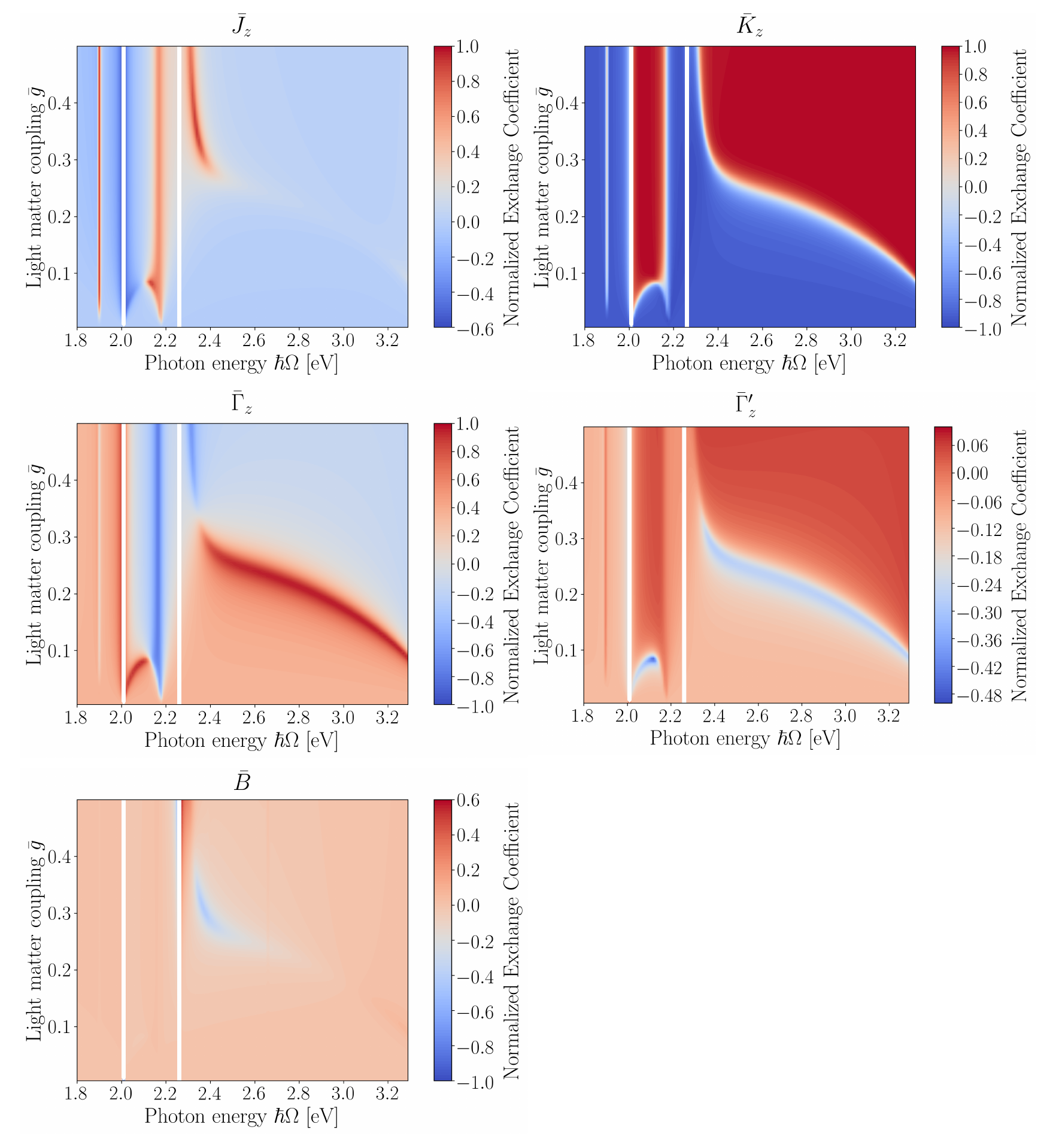}
 \caption{{\bf Spin parameters of the seeded cavity.} Normalized magnetic exchange interaction $\bar{J}$, Kitaev interaction $\bar{K}$, anisotropy interactions $\bar{\Gamma}$ and $\bar{\Gamma}'$ and induced magnetic field $\bar{B}$ as a function of the effective light-matter coupling $\bar{g}$ and the photon energy $\hbar\Omega$ in the zero photon sector $n_{\rm av} = 1$.}
 \label{fig:normalized_seeded_parameters}
\end{figure*}

\end{document}